\documentclass[11pt]{iopart}
\usepackage{bm}
\usepackage{hyperref}
\usepackage{amssymb}
\usepackage{amsopn}

\usepackage{subfigure}
\usepackage{graphicx}

\def\disp{\displaystyle}

\newcommand{\be}{\begin{equation}}
  \newcommand{\ee}{\end{equation}}
\newcommand{\ba}{\begin{eqnarray}}
  \newcommand{\ea}{\end{eqnarray}}

\newcommand{\bmp}{\bar{m}_p}
\newcommand{\ka}{\kappa}
\newcommand{\de}{\delta}
\newcommand{\De}{\Delta}
\newcommand{\h}{{\cal H}}

\newcommand{\adq}{\left | \delta_Q\right |}
\newcommand{\adqi}{\left | \delta_{Q i}\right |}
\newcommand{\ADqi}{\left | \Delta_{Q i}\right |}
\newcommand{\dvdq}{\displaystyle{\frac 1V \frac{\mathrm{d}V}{\mathrm{d}Q}}}
\newcommand{\advdq}{\displaystyle{\left |\frac 1V \frac{\mathrm{d}V}{\mathrm{d}Q}\right |}}

\renewcommand{\(}{\left(}
\renewcommand{\)}{\right)}
\renewcommand{\[}{\left[}
\renewcommand{\]}{\right]}

\renewcommand{\d}{\mathrm{d}}

\begin{document}

%\title[A model of large isocurvature fluctuations in quintessence]
\title{A model of large quintessence isocurvature fluctuations and low CMB quadrupole}

\author{Khamphee Karwan}

%\affiliation{Institut f\"ur Theoretische Physik, Universit\"at Heidelberg,Philosophenweg 16, 69120 Heidelberg, Germany\\
\address{Institut f\"ur Theoretische Physik, Universit\"at Heidelberg,Philosophenweg 16, 69120 Heidelberg, Germany}
\address{
Present Address:
Theoretical High-Energy Physics and Cosmology Group,
Department of Physics, Chulalongkorn University, Bangkok 10330, Thailand
}

\begin{abstract}
Based on quintessence models with anticorrelation between the 
quintessence fluctuations and adiabatic perturbations in the other matter components,
we show that if the quintessence potential is sufficiently steep in the initial era,
i.e., at the beginning of radiation dominated epoch,
the quintessence fluctuations can be large enough to sufficiently suppress
the CMB power spectrum at low multipoles without excess gravitational waves.
The leaping kinetic term quintessence and the crossover quintessence models,
can realize such steep potential at the initial stage,
and therefore can give rise to a low CMB quadrupole as required by observations.

\vspace{3mm}
\begin{flushleft}
\end{flushleft}
\end{abstract}

\maketitle

\section{Introduction}

The power spectrum of CMB temperature fluctuations,
measured by COBE \cite{cobe} and WMAP \cite{wmap:06},
has a smaller amplitude than expected theoretically on large angular scales.
A small amplitude of the CMB fluctuations
on large scales may be a consequence of
the data analyses and limits of observation
\cite{Efstathiou:03, Bielewicz:04, Eriksen:04}.
However, this may also
be due to new physics for
suppressing the CMB power spectrum at low multipoles.
Many models of suppressing the CMB power spectrum
at low multipoles have been proposed
\cite{Contaldi:03, Tsujikawa:03, Piao:05, Wu:06, Campanelli:06}.
One possible model is based on the modification
of the integrated Sachs-Wolfe (ISW) effect by
dark energy perturbations \cite{moroi:03,gorHu:04}.
Dark energy perturbations can suppress the
CMB power spectrum at low multipoles if their amplitude is large 
enough to modify the evolution of metric perturbations
during dark energy domination.
Moreover, they must have anticorrelation
with the curvature perturbations, otherwise they enhance
the low multipoles of the CMB spectrum \cite{kawasaki:01cos}.

In this work, we concentrate on the scalar field model of dark energy,
called quintessence \cite{Wetterich:fm, Ratra:1987rm, caldwell:97}.
We show that if the potential of the quintessence field
is steep enough at the initial stage, i.e., at the beginning of radiation dominated era,
the amplitude of the quintessence fluctuations can be large enough to
enhance the ISW contribution. As a result, the amplitude of the CMB
power spectrum at low multipoles can be suppressed.
We also show that the leaping kinetic term quintessence
and crossover quintessence models can realize such steep potential
at initial stage. 
%
%The enhancement of ISW contribution by dark energy fluctuations is reviewed
%in section 2. In section 3, we show that the quintessence models whose
%potential is steep enough at the end of inflation can give rise to
%large quintessence fluctuations during the present epoch. The suppression of
%the CMB power spectrum at low multipoles by the quintessence fluctuations
%is presented in section 4, and the conclusions are given in section 5, respectively.

\section{The enhancement of ISW contribution by the quintessence fluctuations}

Since the quintessence fluctuations are damped inside the horizon,
we concentrate on the large scales fluctuations.
On large scales, the transfer functions of the CMB temperature fluctuations
can be written in conformal Newtonian gauge as \cite{gorHu:04,hu:95}
\ba
T_\ell(k,\eta_0) &=& {\sqrt{2 \ell(\ell+1)} \over \zeta_i} 
\Big[  {({\Theta_* +\Phi_*}) j_\ell(kD_*)} 
+ \int_{\eta_*}^{\eta_0}\d\eta
 (\Phi' - \Psi')j_\ell(kD) \Big], \nonumber\\
 & \approx & - {\sqrt{2 \ell(\ell+1)} \over \zeta_i} 
\Big[  {1 \over 3}\Psi_{*}  j_\ell(kD_*) 
 + 2\int_{\eta_*}^{\eta_0}\d\eta \,
  \Psi' \, j_\ell(kD) \Big],
\label{trans}
\ea
where a prime denotes a derivative with respect to the conformal time $\eta$,
$\Theta$ is the temperature monopole, $\Phi$ is the gravitational potential,
$\Psi$ is the curvature fluctuation, $j_\ell$ is the spherical Bessel function,
$\zeta_i$ is the initial value of the curvature perturbation on comoving hypersurfaces,
$D = \eta_0-\eta$
and the subscripts $*$ and $0$ denote evaluation at recombination and present respectively.
The angular power spectrum of the CMB can be expressed in terms of the transfer function as
\be
{\ell (\ell+1) C_\ell \over 2\pi} = \int {\d k \over k}
T_\ell^2(k,\eta_0) {\cal P}_{\zeta_i}(k),
\label{cl}
\ee
here ${\cal P}_{\zeta_i}(k)$ is the power spectrum of $\zeta_i$, defined as
\be
%{\cal P}_{\zeta_i} = \frac{k^3}{2\pi^2}\left< \zeta_i^2 \right>.
\left < \zeta_i({\bf k})\zeta_i({\bf k}') \right > =
(2\pi)^3 \delta({\bf k}-{\bf k}') {2\pi^2 \over k^3} {\cal P}_{\zeta_i}(k).
\ee
The first term on the right hand side of eq. (\ref{trans}) corresponds to
the gravitational redshift effects due to the photon's climb out of
the potential well at last scattering. This is the ordinary Sachs Wolfe (SW) effect.
The second term describes the fluctuations induced by the passage of CMB photons through the
time evolving gravitational potential. This is the Integrated  Sachs Wolfe (ISW) effect.
To obtain the second line, we assume that the density fluctuations are adiabatic 
before quintessence domination, and that $\Phi = -\Psi$, i.e. anisotropic stress is negligible.

It follows from eq. (\ref{trans}) that the ISW contribution can cancel
the ordinary Sachs Wolfe contribution if after last scattering $\Psi'$ and $\Psi$ have opposite signs, i.e.
the amplitude of $\Psi$, $|\Psi|$, decreases with time.
Because of the expansion of the universe, the amplitude of $\Psi$ decreases with time,
leading to a partial cancellation of the ISW effects and the ordinary Sachs Wolfe effect.
The contributions from the density perturbations can change the decay rate of $|\Psi |$,
i.e. the ISW contribution.
For quintessence, the quintessence fluctuations can increase the decay rate of $|\Psi |$,
leading to a small amplitude of $\Theta_\ell$ and also $c_\ell$ at low multipoles.
The quintessence fluctuations can influence the evolution of $\Psi$
when quintessence contributes a significant fraction of the total energy density.
We assume that about the end of matter domination radiation can be neglected, so that
the perturbed Einstein equation can be written as \cite{doran:03gip}
\be
\Psi' + \h\Psi + \frac{k^2}{3\h}\Psi 
= \frac 12 \h \(\Omega_m\delta_m + \Omega_Q\delta_Q\), 
\label{eins}
\ee
where the subscripts $m$ and $Q$ denote matter and quintessence, $\Omega$ is the density parameter,
$\de$ is the density contrast, $\h = a' / a$, and $a$ is the scale factor.
The above equation shows that
the quintessence fluctuations can affect the evolution of $\Psi$
if $\adq$ is large enough.
The decay rate of $|\Psi |$ increases when $\de_Q$ and $\Psi$ have opposite signs,
and decreases when $\de_Q$ and $\Psi$ have the same signs,
Hence, the quintessence fluctuations can lead to the
suppression of the CMB power spectrum at low multipoles
if $\adq$ is large enough and
$\de_Q$ and $\Psi$ have opposite signs, i.e. they are anticorrelated.

\section{A model of large quintessence fluctuations}

If the quintessence field $Q$ is light during inflation,
it is nearly frozen and acquires quantum fluctuations \cite{riotto:02}
\be
\delta Q(k)_{\rm inf} = \frac{H_e}{\sqrt{2k^3}},
\label{q_inf}
\ee
where $H_e$ is the Hubble parameter evaluated at the time of
horizon exit, the quintessence field fluctuations $\delta Q$ is in
the conformal Newtonian gauge
and $k$ denotes the wavenumber of the perturbation mode.
When the quintessence field is nearly frozen,
its density contrast $\delta_Q$ in conformal Newtonian gauge is given by
\be
\delta_{Q} = \frac{\delta\rho_{Q}}{\rho_{Q}} \simeq \frac 1{V(Q)} \frac{\d V(Q)}{\d Q}\delta Q,
\label{delta_q}
\ee
where $\rho_{Q}$ is the energy density of quintessence and
$V(Q)$ is the potential of the quintessence field.
As long as the quintessence field is nearly frozen, its fluctuation $\delta Q$
is approximately constant.
Thus if the quintessence field is nearly frozen until the
beginning of the radiation dominated epoch, the initial value of
$\delta_Q$ in the radiation dominated era
is given by
\be
\delta_{Q i} \approx \frac 1V \frac{\d V}{\d Q}\delta Q_{\rm inf}.
\label{delta_qi}
\ee
The evolution of $\de_Q$ during the radiation and matter domination
depends on the evolution of the quintessence field.
Here, we consider the evolution of $\de_Q$ only on superhorizon scales
because $\adq$ rapidly decreases with time on subhorizon scales.
The evolution of the quintessence field is characterized by
the equation of state parameter $w_Q$ which is the ratio of the the pressure to the energy density.
The quintessence fluctuations can be classified into adiabatic and isocurvature modes,
according to the initial conditions.
The adiabatic mode is the mode for which the entropy perturbations vanish,
while the isocurvature modes correspond to the modes for which
the curvature perturbations vanish or approximately vanish \cite{lan:99}.
On large scales, the adiabatic mode is constant \cite{malquarti:02qp,abramo:01}
if $w_Q$ is nearly constant.
Nevertheless, its amplitude is limited by the adiabatic conditions,
so that $\adq$ cannot be adjusted to suppress the CMB power spectrum at low multipoles.
For the isocurvature modes, their evolution depend on the value of $w_Q$.
These perturbations modes decrease rapidly on large scales when $|w_Q|$
is approximately constant and smaller than $1$.
In the context of tracking quintessence,
this corresponds to the case when the quintessence field is in
the tracking regime \cite{stein:98,brax:00}.
To avoid the coincident problem,
the tracking quintessence is in the tracking regime at present.
The expansion of the universe is accelerating today because $w_Q$ in the tracking regime is smaller than $-1/3$.
Before the quintessence field starts the tracking regime,
it is nearly frozen because $w_Q$ is close to $-1$.
The transition from the potential regime to tracking regime occurs between the radiation dominated
and quintessence dominated epochs.
The redshift at which the transition takes place depends on the initial conditions of the quintessence field.
Since $w_Q\approx -1$ when the potential energy of quintessence dominates the kinetic energy,
the regime in which the quintessence field is frozen is called the potential regime.
During this regime, the isocurvature modes are
approximately constant \cite{malquarti:02qp,gorHu:04}.
In general,
the quintessence field can be frozen if the Hubble drag term $3H\dot{Q}$
in its evolution equation dominates the potential term $\d V / \d Q$.
Here, a dot denotes the time derivative and $H$ is the Hubble parameter.
In the case of tracking quintessence,
there may exist a short period of kinetic regime
between inflation and potential regime.
In this regime, the kinetic energy of the quintessence field dominates the potential energy.
The existence of this regime depends on
the choice of the initial conditions for the quintessence field.
The transition from the kinetic regime to potential regime occurs deep in the radiation dominated era.
The isocurvature modes increase during this regime.
Since the kinetic regime is short and
$\adq$ decreases during the transition
from the kinetic regime to the potential regime \cite{malquarti:02qp,brax:00},
the total growth of $\adq$ is small.

We are interested in the suppression of CMB spectrum at low multipoles
via the quintessence fluctuations, which can occur when $\adq$
is large enough during quintessence domination.
From the above summary, we see that the magnitude of $\delta_Q$
can be large during quintessence domination
if its initial value is large and
the quintessence field is nearly frozen initially and remains frozen until quintessence domination.
If the quintessence field is nearly frozen until quintessence domination,
the amplitude of $\de_Q$ during quintessence domination will be approximately equal to
its initial value, given in eq. (\ref{delta_qi}).
Thus, the amplitude of $\de_Q$ during quintessence domination can be large if
the amplitude of $\de Q_{\rm inf}$ or the initial value of
$\advdq$ is large.
A large initial value of $\advdq$ implies a large $\advdq$ near the present epoch
because $\advdq$ is approximately constant as long as the quintessence field is nearly frozen.
The amplitude of $\delta Q_{\rm inf}$ usually depends on $H_e$,
while the initial value of $\advdq$ depends on the quintessence model.
For simple quintessence models, whose potential slope does not change much during quintessence evolution,
the ratio $\advdq$ cannot be large initially
because the quintessence potential must be flat enough at present
to drive an accelerated expansion of the universe today.
Hence, the CMB spectrum at low multipoles can be suppressed sufficiently
if $\delta Q_{\rm inf}$ is large enough.
In the simplest case, this requires $H_e$ to be larger than
the observational bound \cite{gorHu:04}, i.e. there are excess gravitational waves.
This excess gravitational waves problem can be solved 
if the field fluctuation $\delta Q$ is amplified between the
inflationary and present epoch.
This amplification can occur
if the kinetic coefficient of the quintessence field varies in time \cite{gordon:05}.

Alternatively, the problem of excess gravitational waves can be avoided if the
quintessence potential is steep enough initially.
This is because the steep potential can lead to a large $\adqi$
although the amplitude of $\de Q_{\rm inf}$ is small.
However, the quintessence potential must be flat enough at present
to drive an accelerated expansion of the universe.
We now consider whether the quintessence potential can be sufficiently steep initially
and flat enough to drive an accelerated expansion of the universe today.
Let us consider the simple exponential quintessence model, whose potential
can be written as $V(Q) = \bmp^4\exp\(-\lambda Q/\bmp\)$,
where $\bmp = (8\pi G)^{-1/2}$ is the reduced Planck mass.
Thus, $\dvdq = -\disp{\frac \lambda\bmp}$, and
therefore the magnitude of $\de_{Q i}$ can be large if $\lambda$ is large.
In the early epoch, the quintessence field can be frozen although its potential is steep,
because quintessence is not a dominant component.
Using the evolution equation for the quintessence field,
one can show that \cite{stein:98}
\be
\advdq = \frac{\lambda}{\bmp}
= \frac{\sqrt 3}{\bmp\sqrt{\Omega_Q}}\sqrt{1+w_Q}\left | 1+ \frac{1}{6} \frac{d\, \ln{x}}{d\,\ln{a}}\right |,
\ee
where $x=(1+w_Q)/(1-w_Q)$.
The above equation shows that, although $\lambda$ is large, the quintessence field can be frozen,
i.e., $w_Q\approx -1$,
if $\Omega_Q$ is sufficiently small.
In order to drive an accelerated expansion of the universe,
the quintessence field has to be slowly rolling at the present epoch.
When the quintessence field is slowly rolling, its evolution equation yields
\be
\frac 1{3H^2}\frac{d^2V}{dQ^2}\simeq -\frac{\dot H}{H^2}.
\ee
This is a slow-roll condition for the quintessence field.
During the present epoch, quintessence is the dominant component, so that
$\dot H = -\dot{Q}^2/(2\bmp^2)$.
Since $V > \dot{Q}^2/2$and $H^2 \simeq V/(3\bmp^2)$ when the quintessence field is slowly rolling, the slow-roll condition requires
$\lambda^2 < 1.5$, i.e. the quintessence potential should not be too steep.
Unfortunately, the amplitude of CMB quadrupole will be in agreement with the observation
if $\lambda \gg 1$.
However, if $\lambda$ is a time-dependent parameter,
the quintessence potential might be able to be sufficiently steep initially
and flat enough at present.
An exponential quintessence model with time dependent $\lambda$ is 
conveniently described by a leaping kinetic term model \cite{wetterich:00}.
The Lagrangian of this quintessence is
\be
{\cal L}(\chi)=\frac{1}{2}\,(\partial\chi)^2\,\ka^2(\chi )-\bmp^4\exp[-\chi / \bmp]\,.
\label{action}
\ee
Using the field variable $Q=K(\chi)$, where $\ka (\chi)=\partial Q/\partial\chi$, the above Lagrangian becomes
\be
{\cal L}(Q)=\frac{1}{2}\,(\partial Q)^2-\bmp^4\exp[-K^{-1}(Q)/\bmp]\,.
\ee
It is convenient to use the field $Q$ to study the evolution of quintessence because its kinetic term has a canonical form.
The evolution of this field depends on the form of $\ka(\chi )$. We use
\be
\ka(\chi )=\ka_{\rm min}+\mbox{tanh}\(\frac{\beta}{\bmp}\[\chi - \chi_1\]\)+1,
\label{3kinc}
\ee
where $\ka_{\rm min}$, $\chi_1$ and $\beta$ are constant.
In 			the early epoch, $\chi\ll\chi_1$ so $\ka\simeq\ka_{\rm min}$.
Therefore, we have
$\dvdq\simeq -(\bmp\ka_{\rm min})^{-1}$.
In our consideration, the field $\chi$ and $Q$ are nearly frozen during the initial stage.
As the universe evolves, the Hubble drag term decreases due to decreasing Hubble parameter,
while the potential term is approximately constant because $Q$ and $\chi$ are nearly frozen.
When the Hubble drag term is small enough,
the fields $Q$ and $\chi$ are able to roll down their potentials.
Consequently, the field $\chi$ increases and
becomes larger than $\chi_1$ during matter domination.
As a result, the kinetic coefficient $\ka$ increases and
therefore $w_Q$ decreases towards $-1$ again.
The evolution of this quintessence field
has a tracking behavior.
The present value of $w_Q$ depends on $\beta$,
while the present value of $\Omega_Q$ depends on $\chi_1$ if $\beta$ is fixed.
The evolution of $w_Q$ is shown in figure 1.

\begin{figure}[ht]
\begin{center}
\includegraphics[height=0.4\textwidth, width=0.8\textwidth,angle=0]{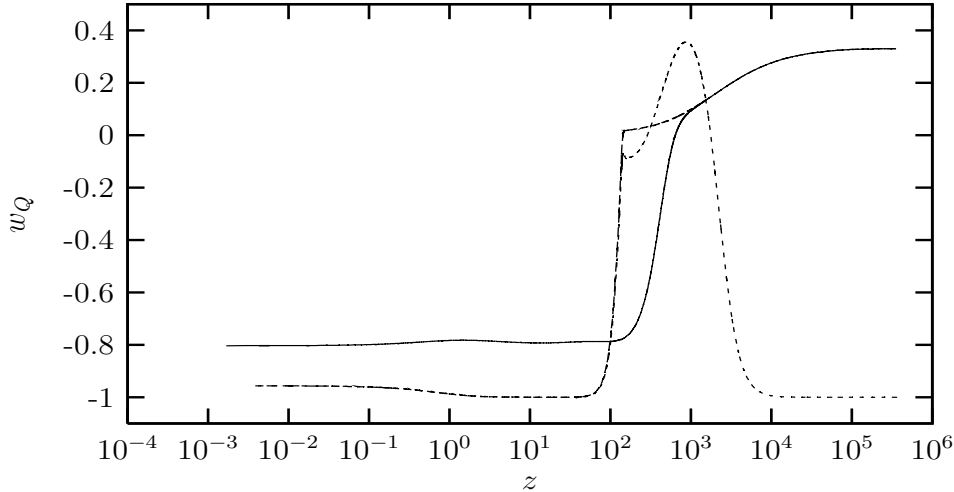}
\caption{
Evolution of $w_Q$. The solid line and long dashed line show
the tracking solutions, where $\beta=1$ for the solid line and $\beta=500$
for the long dashed line. The dashed line corresponds to the case when the quintessence
field starts in the potential regime and $\beta=500$. In this plot,
we set $\ka_{\rm min}=8.7\times 10^{-4}$
and $\Omega_{Q}=0.7$ at the present epoch.
}
\label{fig:wq}
\end{center}
\end{figure} 

It is easy to see that $\adqi$ is large when $\ka_{\rm min}$ is small.
At first sight, this looks similar to the existing literature because the magnitude
of the quintessence fluctuations is enhanced by a
 kinetic coefficient \cite{gordon:05}.
Nevertheless, the physical motivation is different.
In the literature, the quintessence field fluctuation is
amplified between the inflationary and present epoch
by a varying kinetic coefficient.
In our analysis, the exponential quintessence can give rise to a large
$\adqi$ if $\lambda$ is large. The varying kinetic coefficient is used to push
$w_Q$ towards $-1$ at the present epoch
without amplifying the fluctuations in the quintessence field.
The high energy physics model of this type of quintessence, called crossover quintessence,
has been proposed by Wetterich \cite{wetterich:02}.
After suitable redefinition of dynamical variables,
the action of crossover quintessence takes the form as shown in eq. (\ref{action}).
The evolution of crossover quintessence is not exactly the same as
the evolution of leaping kinetic term quintessence because their kinetic coefficients have different forms.
However, both quintessence models evolve in the same manner, i.e., they evolve in the early epoch
as a simple exponential quintessence model and their equation of state parameter is pushed
towards $-1$ at present by rapid increase of the kinetic coefficient.
Since the form of the kinetic coefficient of crossover quintessence is more
complicated than that of the leaping kinetic term quintessence,
we use a leaping kinetic term quintessence model in our consideration.

\section{The suppression of the CMB power spectrum}

\begin{figure}[ht]
\begin{center}
\includegraphics[height=0.4\textwidth, width=0.8\textwidth,angle=0]{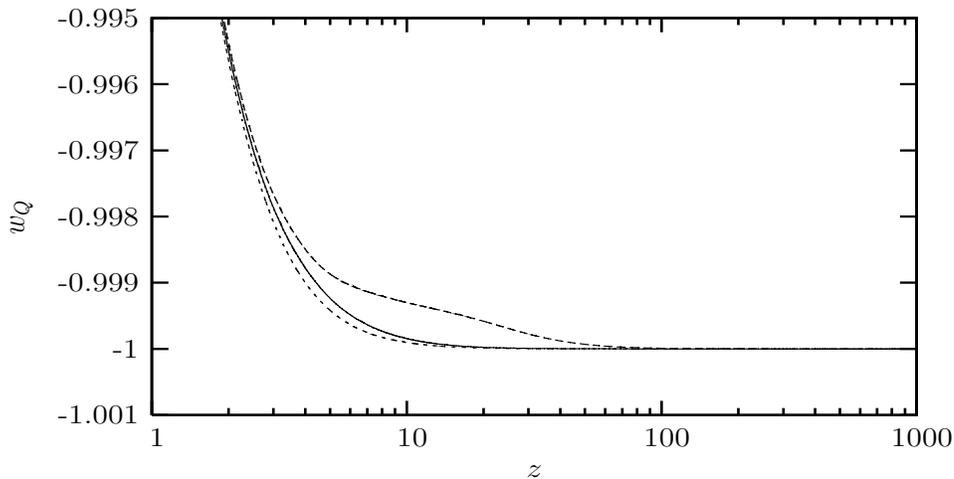}

\caption{
Evolution of $w_Q$.
The long dashed line represents the case where
the quintessence fluctuations have no effect on the evolution of $\Psi$,
the dashed line represents the case where
the quintessence fluctuations lead to a large ISW contribution and
the solid line represents the case where
the CMB power spectrum at low multipoles is sufficiently
suppressed by the quintessence fluctuations.
For these line styles, we set $\ka_{\rm min}=8.7\times 10^{-4}$
and $\beta = 500$.
The value of $Q_i$ for the long dashed lines is smallest,
while the one for the dashed lines is largest.
These representations of line styles will be used in all subsequent figures.
}
\label{fig:wslow}
\end{center}
\end{figure}

We first consider the initial conditions for the quintessence field and
the quintessence fluctuations.
Since we suppose that the quintessence field $Q$
is frozen initially, it will be
frozen until quintessence domination if its initial value $Q_i$
is larger than the tracking solution value \cite{stein:98}.
In our consideration, the value of $Q_i$ is chosen such that
$w_Q$ evolves as shown in figure \ref{fig:wslow}.
It can be seen in this figure that the
redshift $z_c$ at which
the quintessence field leaves the potential regime decreases
if $Q_i$ increases.
The evolution of $\Omega_Q$ is approximately the same for
all chosen $Q_i$ so we do not plot it.

To compute the initial conditions for the quintessence fluctuations,
we suppose that the isocurvature fluctuations in quintessence have
(anti-) correlation with adiabatic fluctuations in the other matter,
so that the initial conditions for quintessence fluctuations can be written
as \cite{gorHu:04} $\de_{Q \rm int} = \de_{Q \rm ad} + \de_{Q \rm iso}$
and $u_{Q \rm int} = u_{Q \rm ad} + u_{Q \rm iso}$.
Here, the subscripts int, ad and iso denotes
the mixed initial conditions, adiabatic initial conditions
and isocurvature initial conditions respectively.
The density contrast $\de_Q$ and momentum density $u_Q$
in these expressions are in conformal Newtonian gauge.
When the quintessence field is nearly frozen,
the density contrast $\de_Q$
is roughly equivalent to the
gauge invariant density contrast $\De_Q$
because $\De_Q = \de_Q + 3(1+w_Q)\Psi$.
Moreover, the momentum density $u_Q$ and
the gauge invariant momentum density $U_Q$
are the same. Thus,
we can use the gauge invariant formulas in \cite{doran:03gip}
to compute the adiabatic initial conditions for quintessence
and the other species, e.g., photon,
baryon, neutrino and CDM.
These initial conditions are written in terms of $\zeta_i$,
where $\zeta_i=1$.
For isocurvature modes,
the value of $\De_{Q \rm iso}$ is computed
using eq. (\ref{delta_qi}).
The field fluctuation $\de Q_{\rm inf}$
is written
in terms of curvature perturbation as \cite{riotto:02,wmapInf:03}
$\de Q_{\rm inf} \simeq \sqrt{2\epsilon}\zeta_i\bmp\simeq\sqrt{0.125 R}\zeta_i\bmp$,
where $\epsilon$ is the slow roll parameter of inflaton
and $R$ is the relative amplitude  of the tensor to  scalar perturbations.
If we hold $R$ fixed,
the value of $\De_{Q \rm iso}$ will depend only on $\ka_{\rm min}$.
Since the observational bound of
$R$ is $R<0.28$
at the 95\% confidence level \cite{wmap:06},
we set $R=0.01$ in our calculation.
The choice of $R$ does not affect the results of the calculation
because different choices of $R$ can give the same results
if the value of $\ka_{\rm min}$ is adjusted.
The value of $U_{Q \rm iso}$ can be computed
using \cite{gorHu:04} $\De_{Q \rm iso}= U_{Q \rm iso}\(s - 3w_Q + 3c_{a Q}^2 -1\)$,
where
$s=\frac 12 \[6w_Q-3c_{a Q}^2-2 \pm \sqrt{9c_{a Q}^4+12c_{a Q}^2-20}\]$
and $c_{a Q}^2$ is the adiabatic sound speed of quintessence.
In our case, the quintessence field is frozen initially so
$c_{a Q}^2=-7/3$ \cite{brax:00} and
therefore $U_{Q \rm iso}=-\De_{Q \rm iso}/5$.

\begin{figure}[ht]
\begin{center}
\includegraphics[height=0.4\textwidth,width=0.45\textwidth,angle=0]{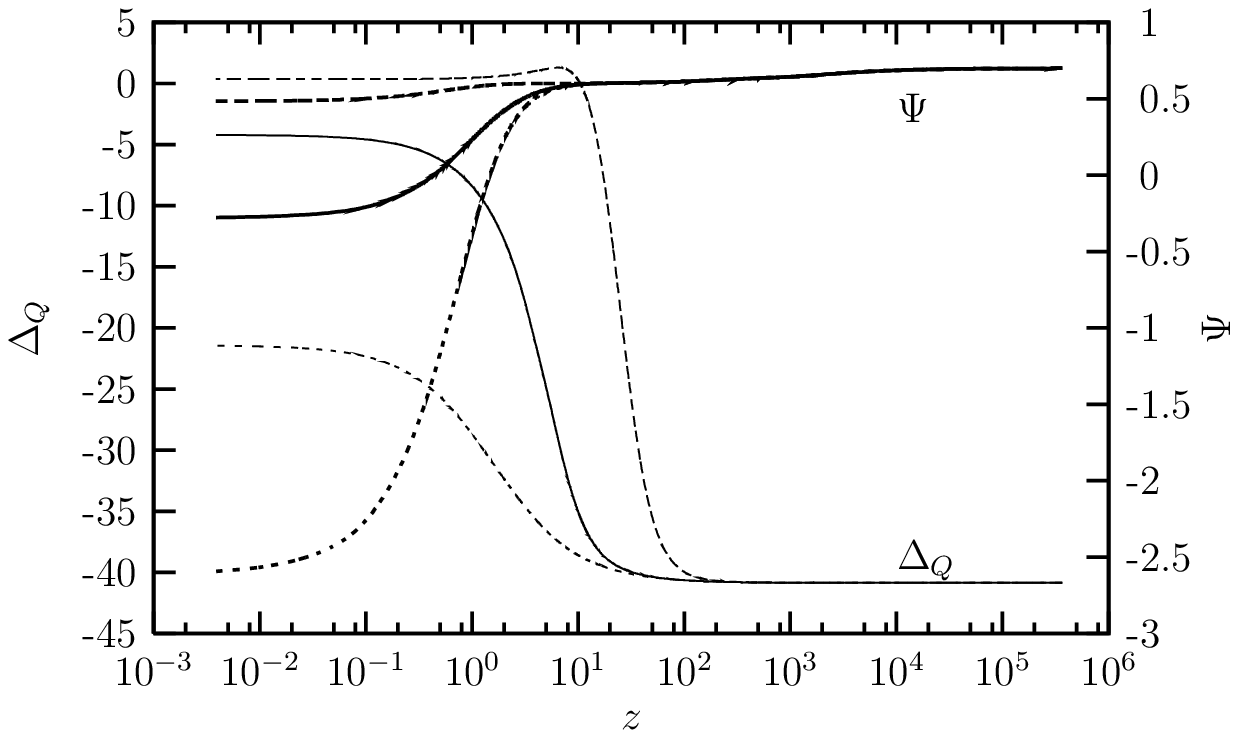}
\includegraphics[height=0.4\textwidth,width=0.45\textwidth,angle=0]{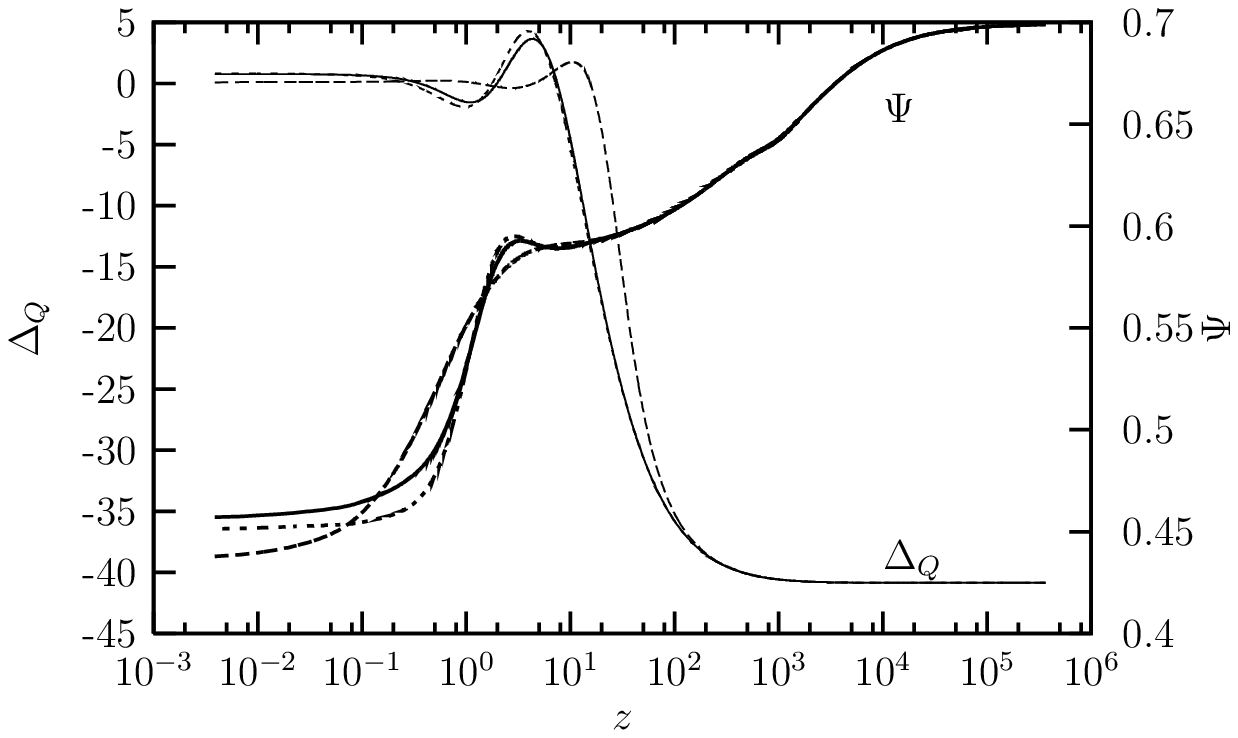}

\caption{
Evolution of $\De_Q$ (thin lines) and $\Psi$ (thick lines)
for the models whose $w_Q$ is plotted in figure \ref{fig:wslow}.
The left panel shows a mode whose wavelength is larger
than the horizon today ($k=2\times 10^{-4} \mbox{Mpc}^{-1}$),
while the right panel shows a mode which enters the horizon about
the end of matter domination ($k=1\times 10^{-3} \mbox{Mpc}^{-1}$).
}
\label{fig:delq}
\end{center}
\end{figure} 

We use CMBEASY \cite{cmbeasy} to compute the evolutions of $\De_Q$ and $\Psi$
and plot them in figure \ref{fig:delq}.
From this figure, we see that the magnitude of $\De_Q$
decreases when $w_Q$ increases from $-1$.
For a given redshift, the decay rate of $\Psi$
increases with increasing $\left | \De_Q \right |$.
Thus, the ISW contribution increases with decreasing $z_c$ (or equivalently with increasing $Q_i$).
This is because $\left | \De_Q \right |$ will start to decrease at lower redshift
if $z_c$ decreases.
On small scales, the quintessence fluctuations decrease
after horizon crossing.
Thus, the quintessence fluctuations lead to
less modifications of $\Psi'$ compared with the case of large scales.
Because of the oscillation of the quintessence fluctuations,
$\Psi'$ oscillates after horizon crossing.
Therefore, on small scales, the amount
of the ISW contribution does not increase with decreasing $z_c$.

\begin{figure}[ht]
%\begin{center}
\includegraphics[width=0.8\textwidth,angle=0]{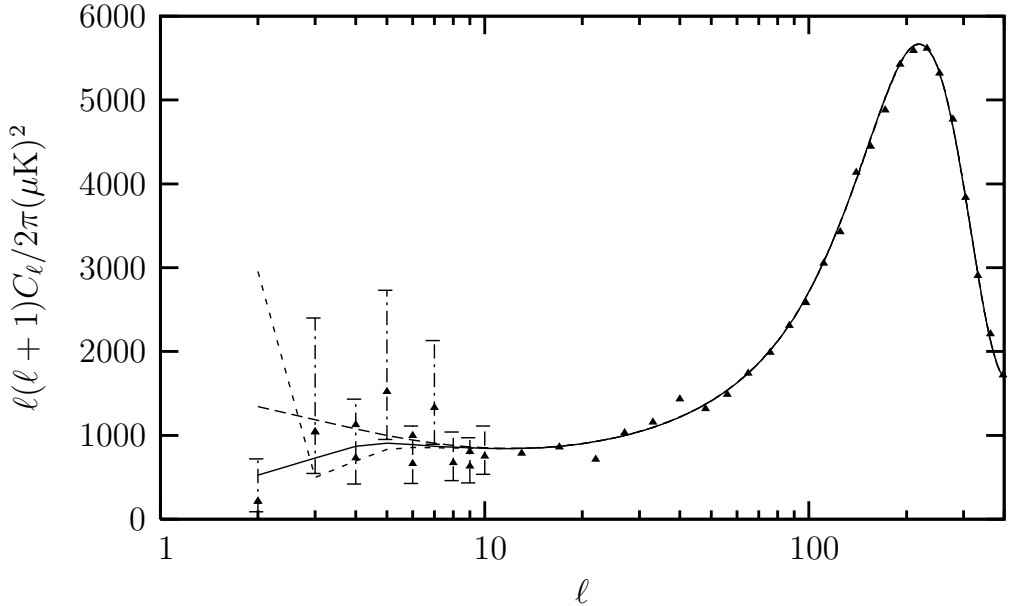}
\begin{center}
\caption{
The CMB power spectrum for different choices of $Q_i$.
The data points from WMAP 3-year results are denoted by the solid triangles.
The error bars show the 68$\%$ confidence level.
}
\label{fig:cmb}
\end{center}
\end{figure} 

The CMB power spectrum is computed using CMBEASY and
shown in figure \ref{fig:cmb}.
In this figure,
we set $\Omega_bh^2=0.0223$,
$\Omega_mh^2=0.127$,
$\Omega_Q=0.73$,
$h=0.73$,
$\tau=0.09$,
$n_s=0.95$ and
$\beta=500$.

From this figure,
we see that the amplitude  of the CMB power spectrum at low multipoles
decreases with decreasing $z_c$ for a suitable range of $z_c$.
This is because the ISW contribution increases when $z_c$ decreases.
The value of $z_c$ must
be smaller than the redshift of matter-radiation equality,
otherwise the amount of the quintessence fluctuation 
is not large enough to modify the ISW contribution.
Nevertheless, if $z_c$ is too small,
the CMB quadrupole increases but a few higher multipoles decrease
when $z_c$ decreases

Since the ISW contribution depends on the amount of the quintessence fluctuations at low redshift,
the amplitude of the CMB power spectrum at low multipoles also depends on $\ka_{\rm min}$ and $\beta$.
For appropriate values of $z_c$ and $\beta$, the amplitude of the CMB spectrum decreases
with decreasing $\ka_{\rm min}$ due to the increasing $\ADqi$.
The different values of $\ka_{\rm min}$
can lead to the same amplitude of the CMB power spectrum
if the value of $z_c$ for the case of small $\ka_{\rm min}$ is larger
than the one for the case of large $\ka_{\rm min}$.
However, the value of $\ka_{\rm min}$ must not
be too large or too small compared with the value which is chosen here.
If $\ka_{\rm min}$ is too large, the quintessence fluctuation will have
no effect on the CMB power spectrum because $\ADqi$ is too small.
If $\ka_{\rm min}$is too small, the CMB power spectrum at low multipoles
will be enhanced by the quintessence fluctuations due to the large ISW contribution.
Furthermore, for appropriate values of $z_c$, $\ka_{\rm min}$  and a suitable range of $\beta$,
the value of $w_Q$ in the tracking regime gets closer to $-1$ as $\beta$ increases.
Since the amplitude of $\De_Q$ in the tracking regime decreases slower when $w_Q$ is closer to $-1$,
the amplitude of $\De_Q$ at particular redshift in this regime increases with increasing $\beta$.
Hence, the amplitude of CMB power spectrum at low multipoles decreases with $\beta$.

\begin{figure}[ht]
\begin{center}
\includegraphics[height=0.4\textwidth,width=0.45\textwidth,angle=0]{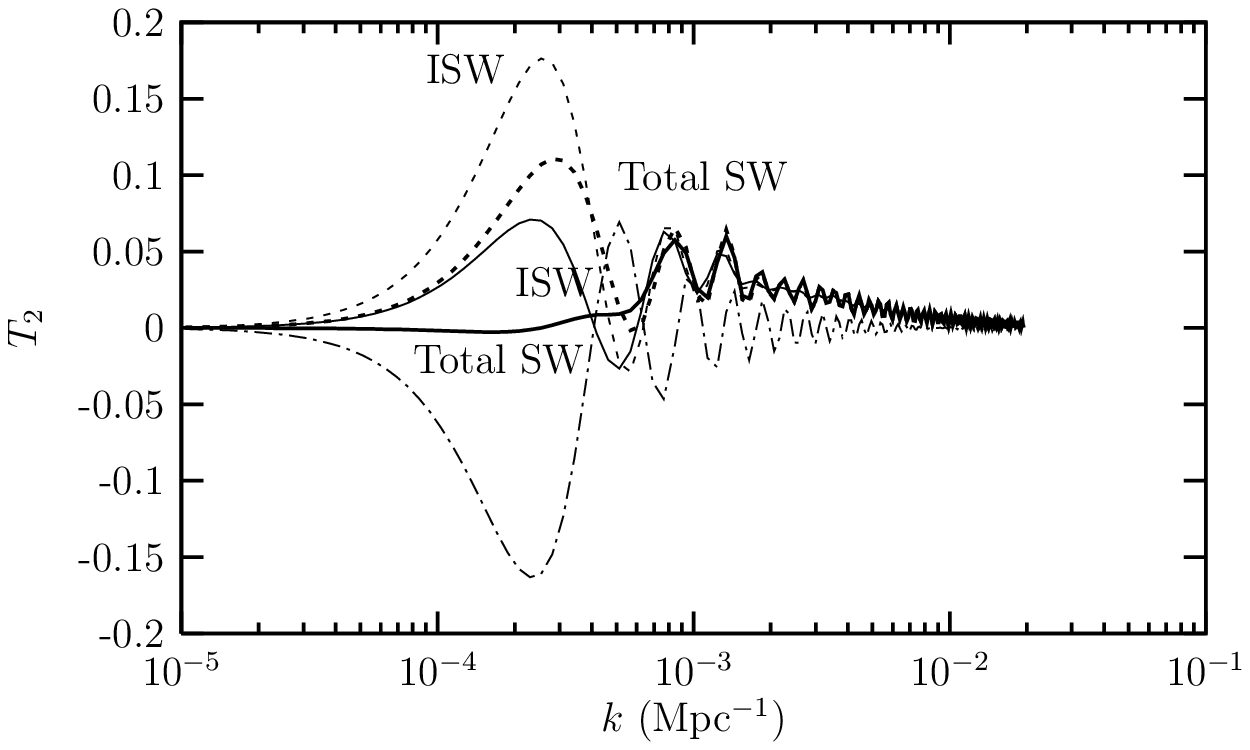}
\includegraphics[height=0.4\textwidth,width=0.45\textwidth,angle=0]{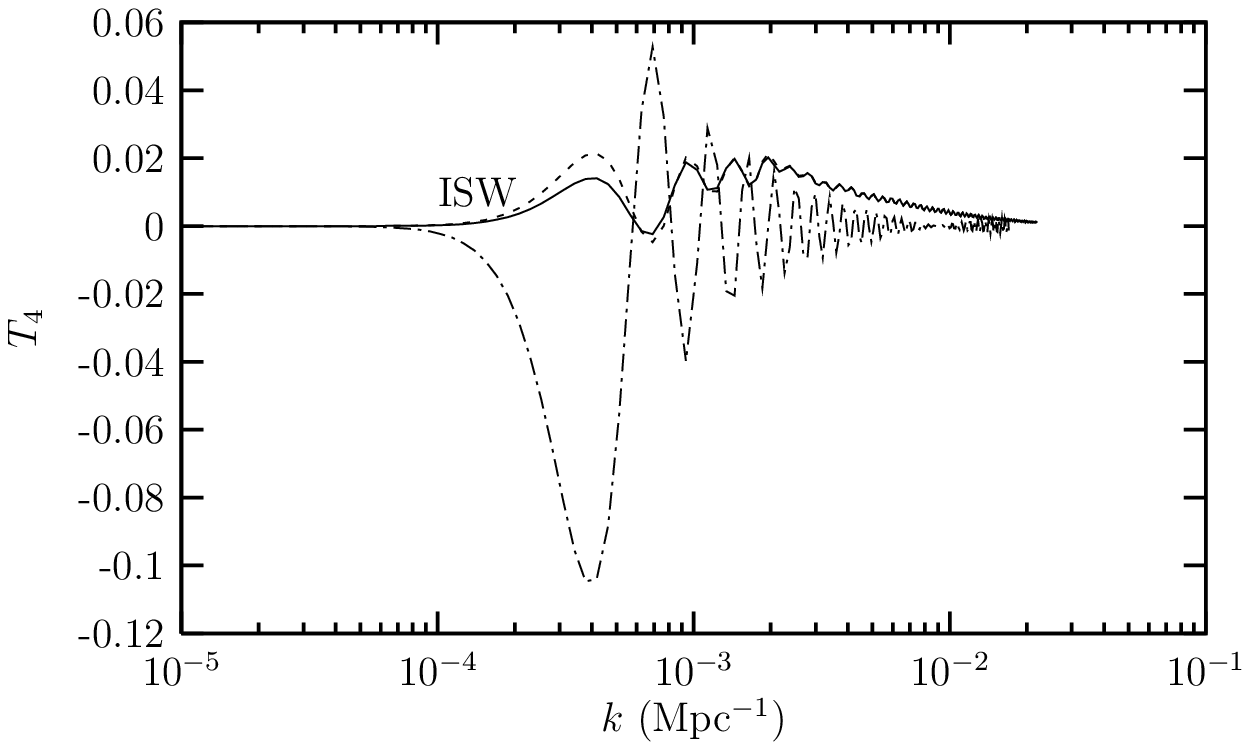}
\caption{
Radiation transfer function of the CMB 
power spectrum in figure \ref{fig:cmb}.
The left panel corresponds to $\ell = 2$,
while the right panel corresponds to $\ell = 4$.
The ordinary SW effect is represented by the dash-dotted lines.
The thin lines represent the ISW effect
and the thick lines represent the total
(ordinary $+$ integrate) SW effect.
The solid and dashed lines correspond to the different ISW contributions
which lead to the CMB power spectra in figure \ref{fig:cmb}.
}
\label{fig:sw}
\end{center}
\end{figure}

The enhancement of the
CMB quadrupole when $z_c$ is too small
can be understood by considering the CMB transfer function.
we plot the CMB transfer functions in figure \ref{fig:sw}.
This figure shows that the ISW contribution for $\ell = 2$ is
much enhanced around its peak due to
the quintessence fluctuations.
If the amount of ISW contribution is too large, it leads
to an enhancement of the CMB quadrupole.
For $\ell = 4$, quintessence fluctuations lead to
 less ISW contribution
because the ISW contribution peaks at a
scale which is smaller than the horizon size today and
the quintessence fluctuations decrease inside the horizon.
Therefore,when $z_c$ is too small,
the CMB quadrupole increases with decreasing $z_c$
although  the CMB spectrum at $\ell = 4$ still decreases with decreasing $z_c$.

\section{Conclusions}

The quintessence fluctuation can lead to
the suppression of the CMB power spectrum at low multipoles
if its amplitude is large enough during quintessence domination and it has
an anticorrelation with the adiabatic fluctuations in the other matter.
Assuming that the quintessence fluctuation is generated
during inflation, its amplitude can be large during quintessence domination if
the amplitude of its initial value $\De_{Q i}\simeq \dvdq\de Q_{\rm inf}$ is large and
the quintessence field is nearly frozen until quintessence domination.
It has been shown that the quintessence field may leave the potential regime
at low redshift because its coarse-grained part is driven towards a large value
by its quantum fluctuations \cite{Malquarti:02si, Martin:05}.
Thus, if the initial amplitude of the quintessence fluctuation is large,
the amplitude of quintessence fluctuations can be large enough to enhance the ISW contribution,
and therefore suppress the CMB power spectrum at low multipoles.
For simple quintessence models, $\left | \De_{Q i}\right |$ can be large
if $\de Q_{\rm inf}$ is large, leading to excess gravitational waves.
In this work, we avoid the excess gravitational waves by
keeping $\de Q_{\rm inf}$ below observational bound.
The large  $\left | \De_{Q i}\right |$ can be obtained by enhancing
the initial value of $\advdq$.
However, To drive an accelerated expansion of the Universe,
the quintessence potential must be flat today.
Hence, the quintessence field usually has small
$\advdq$ in the early epoch.
We found that the potential of leaping kinetic term quintessence
can be sufficiently steep initially and be nearly flat today.
Thus, this quintessence model can produce
large fluctuations to explain the observed low CMB quadrupole
if the value of $\ka_{\rm min}$ is chosen appropriately.
This quintessence model has a tracking behaviour
but the amplitude of the CMB power spectrum at low multipoles
depends on $z_c$, or equivalently
the initial value of the quintessence field.\\

Finally, we note that one of the basic assumption in this work is that
the quintessence fluctuations and the adiabatic perturbations in the other matter must have an anticorrelation.
The genaral idea for genarating the correlated fluctuations has been considered in
\cite{lan:99inf,amen:02}.
A specific model for genarating correlated fluctuations in particular quintessence
model is proposed in \cite{moroi:03}.
Here, we just suppose that the quintessence fluctuations and the
adiabatic fluctuations in the other matter can have an anticorrelation.
We do not propose a specific model for genarating the anticorrelated fluctuations
in the considered quintessence model

\ack
This work is a part of my Ph.D. thesis at University of Heidelberg.
I would like to thank C. Wetterich who is my thesis advisor.
I would also like to thank M. Doran and A. Ungkitchanukit for comments on the manuscript.
I am also grateful for financial support from DAAD (German Academic Exchange Service)
%Deutscher Akademischer Austauschdienst
during my Ph.D. study.

%\begin{thebibliography}{}
\section*{References}

\providecommand{\href}[2]{#2}\begingroup\raggedright\endgroup
 
\end{document}